\documentclass{ecai} 

\usepackage{latexsym}
\usepackage{amssymb}
\usepackage{amsmath}
\usepackage{amsthm}
\usepackage{booktabs}
\usepackage{enumitem}
\usepackage{graphicx}
\usepackage{color}

\newcommand{\BibTeX}{B\kern-.05em{\sc i\kern-.025em b}\kern-.08em\TeX}

\begin{document}


\begin{frontmatter}


\paperid{0920} 


\title{Detecting Struggling Student Programmers using Proficiency Taxonomies}


\author{\fnms{Noga} \snm{Schwartz}\textsuperscript{1},
        \fnms{Roy} \snm{Fairstein}\textsuperscript{1}, 
        \fnms{Avi} \snm{Segal}\textsuperscript{1}, 
        \fnms{Kobi} \snm{Gal}\textsuperscript{1,2}}

\address{\textsuperscript{1}Department of Software and Information Systems Engineering, Ben-Gurion University, Israel\\
\textsuperscript{2}School of Informatics, University of Edinburgh, U.K.}


\begin{abstract}
Early detection of struggling student programmers is crucial for providing them with personalized support. 
While multiple AI-based approaches have been proposed for this problem, they do not explicitly reason about students' programming skills in the model. 
This study addresses this gap by developing in collaboration with educators a taxonomy of proficiencies that categorizes how students solve coding tasks and is embedded in the detection model.
Our model, termed the Proficiency Taxonomy Model (PTM), simultaneously learns the student's coding skills based on their coding history and predicts whether they will struggle on a new task. 
We extensively evaluated the effectiveness of the PTM model on two separate datasets from introductory Java and Python courses for beginner programmers. 
Experimental results demonstrate that PTM outperforms state-of-the-art models in predicting struggling students. 
The paper showcases the potential of combining structured insights from teachers for early identification of those needing assistance in learning to code.

\end{abstract}

\end{frontmatter}


\section{Introduction}
Detecting struggling students is essential for enhancing learning outcomes and reducing dropout rates in education \cite{de2013critical}. Learning to program is inherently complex, requiring students to develop problem-solving and logical reasoning skills \cite{islam2019study, qian2017students}. 

Many students experience persistent difficulties that hinder their progress, yet teachers often lack the tools to detect those in need of additional support. Without early detection, these students may become discouraged, leading to reduced engagement and a higher likelihood of attrition in programming courses \cite{tek2018implicit, gorson2020cs1}. 

To address this challenge, this paper proposes an AI-based approach that combines language model representations of students' coding histories with their inferred programming skills.  Our approach also incorporates qualitative insights from educators through in-depth interviews. These interviews guided the development of a new proficiency taxonomy of students' coding behavior, which aligns educational goals with specific coding skills for students. 
This taxonomy is based in part on Bloom's taxonomy and its adaptations to programming~\cite{bloom1956taxonomy,selby2015relationships,thompson2008bloom}.

Previous research for detecting struggling students using AI has primarily focused on limited historical data, often considering only a subset of past coding attempts rather than a student's entire submission history \cite{tsabari2023predicting, shi2022code, charitsis2022using}. Our approach is able to utilize a student's complete coding history, encompassing all previously attempted code submissions for each task. 

Additionally, while past models predominantly relied on Abstract Syntax Trees (AST) for code representation \cite{shi2022code, shi2024evaluating, yu2024eckt}, our method integrates CodeBERT \cite{feng2020codebert} to generate deep learning-based representations of student code. 
By capturing the full trajectory of student learning, our approach offers a detailed view of student progress.

Our proposed model, termed Proficiency Taxonomy Model (PTM), directly embeds students' coding proficiency skills, ensuring that the predictions are in line with the educators' understanding of struggling students.
It is a multi-task model that simultaneously learns the coding proficiencies for a target student and whether the student will struggle on the next task. 
It uses a separate loss function for each task. 

We apply PTM to detect struggling students in two commonly used student programming datasets, one in Python and the other in Java. 
The datasets differ in task diversity, student participation, and problem complexity, providing a comprehensive evaluation across varying programming environments.
We compare the performance of PTM against commonly used methods in the literature, including Deep Knowledge Tracing (DKT) \cite{piech2015deep}, the Self-Attentive Knowledge Tracing model (SAKT) \cite{pandey2019self}, and Code-DKT \cite{shi2022code}.

For both datasets, our model outperforms these approaches in predicting struggling students, achieving superior performance in terms of ROC-AUC scores. 
We provide an analysis of the performance differences across the datasets, as well as the sensitivity of the model to a varying number of past coding tasks.
Through an ablation study, we confirm that including both coding history representations and students' proficiency taxonomies contribute to the model performance. 

The contributions of this work are in 1) 
the combination of pedagogical insights (students' proficiency taxonomies) in language models of students' programming, 2) a new multi-task detection model that simultaneously learns students' programming skills as well as whether they will struggle on new tasks, 3) extensive empirical evaluation on two different programming environments. 

Our approach highlights the importance of combining input from instructors with the power of language-based models in identifying struggling students. 
Our implementation is publicly available on GitHub~\cite{schwartz2025ptm_github}.

\section{Related Work}
Our work relates to several research areas in AI and education: 
(1) Representing students' code and submissions; 
(2) Detecting struggling students learning to code; and (3) Taxonomies of students' coding proficiencies. We elaborate on each of these areas below.

\subsection{Representing Students' Code Submissions}
Early work explored representing code submissions through Code2Vec, introduced by Alon et al. \cite{alon2019code2vec}, which uses Abstract Syntax Tree (AST) paths to create vector representations of code snippets. 

The introduction of CodeBERT \cite{feng2020codebert} marked a significant advancement by leveraging both programming and natural language data to produce embeddings that capture syntax and semantics, establishing it as a robust foundation for code understanding tasks.

The effectiveness of CodeBERT in various contexts has been further validated by recent studies. Zhao et al.~\cite{zhao2025empirical} compared various code representation models, including newer architectures, and found that CodeBERT provides reliable performance for software programming classification tasks such as code vulnerability detection, code clone detection, and just-in-time defect prediction, particularly when focusing on the vector representation of each code token.
Additionally, Zhang et al.~\cite{zhang2023survey} demonstrated that CodeBERT-based representations can effectively capture structural patterns in student code, serving as a viable component in software programming modeling tasks.

We use CodeBERT for code representation and combine it with LLM-based task text embeddings. Despite newer models, CodeBERT offers a strong balance between performance and practical deployment in education, effectively capturing both syntax and semantics.

\subsection{Detecting Struggling Student Programmers}
Multiple definitions exist in the literature for struggling students.  Alzahrani et al. \cite{alzahrani2018python} and Edwards et al. \cite{edwards2019can} define struggling students based on metadata such as time spent on tasks, number of attempts and success in solving the task. 
Other researchers, such as Tabarsi et al. \cite{tabarsicatch}, Tek et al. \cite{tek2018implicit}, and Gorson and O'Rourke \cite{gorson2020cs1}, define struggling students through behavioral patterns and psychological factors, including self-efficacy, debugging behaviors, and responses to time pressure and error resolution.
In this research, we consider the number of submission attempts and a failure to submit a correct solution as indicators of a struggling student. This is also the approach taken in the CSEDM 2021 competition~\cite{csedm2021competition,penmetsa2021investigate}. 

Identifying struggling students relates to past work in knowledge tracing for predicting student's performance (although the definition of struggling students depends also on attempts, not just performance). 
Deep Knowledge Tracing (DKT) \cite{piech2015deep} uses recurrent neural networks (and specifically long short-term memory networks - LSTMs) to predict the correctness of each problem attempt based on their previous interactions. Shi et al. \cite{shi2022code} introduced Code-DKT, an improved knowledge tracing model that leverages code2vec \cite{alon2019code2vec} representations and task sequences in the prediction task. Shi et al. \cite{shi2024evaluating} introduced an explicit knowledge component (KC) layer to the DKT that leverages an expert-defined Q-matrix to align problems with their respective skills. This approach enhanced the interpretability of DKT and CodeDKT but did not exhibit better performance. 

Our work extends prior knowledge tracing models by combining students' coding proficiencies with their complete coding history and model-based representations of coding tasks.

\subsection{Taxonomies in Programming Education}
Taxonomies offer structured frameworks for categorizing skills and identifying learning challenges. Several have been adapted to programming education.

Bloom's taxonomy \cite{bloom1956taxonomy} classifies cognitive skills into hierarchical levels, later revised to Remember, Understand, Apply, Analyze, Evaluate, and Create \cite{krathwohl2002revision}. It has been applied to computer programming tasks as well.

Selby et al.~\cite{selby2015relationships} linked computational thinking and Bloom's taxonomy, identifying decomposition as particularly challenging. Thompson et al.~\cite{thompson2008bloom} noted difficulties in consistently applying Bloom's taxonomy to programming due to varying teaching contexts.

Building on these works, our Coding Proficiency Taxonomy targets programming education specifically. It adopts Bloom's hierarchical structure to organize student behaviors from reading comprehension to advanced skills like decomposition, edge case handling, and testing. The Coding Proficiency Taxonomy is directly embedded in the model, improving its ability to detect struggling students.

\section{Research Questions and Data} 
\label{sec:research}
Our approach focuses on predicting whether a student will struggle with a new coding task based on their coding history, which includes coding snapshots of all their submissions for past coding tasks. 

Our research questions are as follows:
(\textbf{RQ1}) Can code-based deep language models enhance the accuracy of identifying students who are likely to struggle across multiple datasets?
(\textbf{RQ2}) Does incorporating prior snapshots of students' code improve the ability to identify struggling students when evaluated on multiple programming languages?
(\textbf{RQ3}) Can incorporating a taxonomy of students' coding proficiency enhance prediction performance in Java and in Python?

The study used two freely available datasets of students learning to program. These datasets provide Python and Java programming tasks targeting different programming concepts. Students were allowed multiple submissions for each task and their submissions were automatically graded based on their performance on specified unit tests.

Each task includes instructions, programming concepts, and submission history. The datasets originally contained 18 and 20 concepts, respectively, which we merged into ten: loops, conditional clauses, math operations, logic operations, string manipulations, lists, file operations, functions, dictionaries, and tuples. This enables general analysis across Java and Python (e.g., all loop types defined as “loops”). 

The first dataset comes from the CodeWorkout environment~\cite{edwards2017codeworkout}, which includes submissions of students' solutions to Java programming tasks in an introductory Computer Science course at a public university in the U.S. Students worked on 50 coding tasks, each requiring 10–26 lines of code. We analyzed 630 students in this dataset.

An example of a coding task in CodeWorkout was:
\begin{quote}
    Write a function in Java that implements the following logic: Given a string \texttt{str} and a non-empty word, return a version of the original string where all characters have been replaced by pluses (+), except for appearances of the word which are preserved unchanged.
\end{quote}
The programming concepts targeted by this task were conditionals, loops, and string manipulations.

The second dataset, called FalconCode, contains Python programming tasks from courses taught at the United States Air Force Academy~\cite{de2023falconcode}. We selected introductory programming tasks focusing on fundamental skills (1-3 lines of code) and labs (10–50 lines of code). 
We analyzed 1,330 students and 408 coding tasks in this dataset. Unlike CodeWorkout, FalconCode exhibits wide variation in how many students submit each task, and students typically complete different subsets of the available tasks.

An example of a coding task from this dataset is: 
\begin{quote}
Write an algorithm that asks the user for a positive number and then outputs the multiplication table of that number(from 1-10).
\end{quote}
The programming concepts targeted by this task were string operations, loops, and math operations.

Following past work~\cite{csedm2021competition, penmetsa2021investigate}, a student is defined as struggling if they either
fail one or more unit tests for a given task or require significantly
more attempts than 75\% of their peers to complete
the task successfully.

This definition avoids relying on first-attempt success, as not all students solve the tasks on their first attempt. Tasks that encourage iterative development can increase the number of attempts for all students, making a relative threshold necessary.
Additionally, passing all unit tests is not used on its own as most students eventually succeed in given tasks. 
To illustrate, 179 students out of 630 students who attempted the CodeWorkout example above were defined as struggling; 48 students out of 168 students who attempted the FalconCode example above were defined as struggling.

\begin{figure}
\centerline{\includegraphics[width=7.5cm]{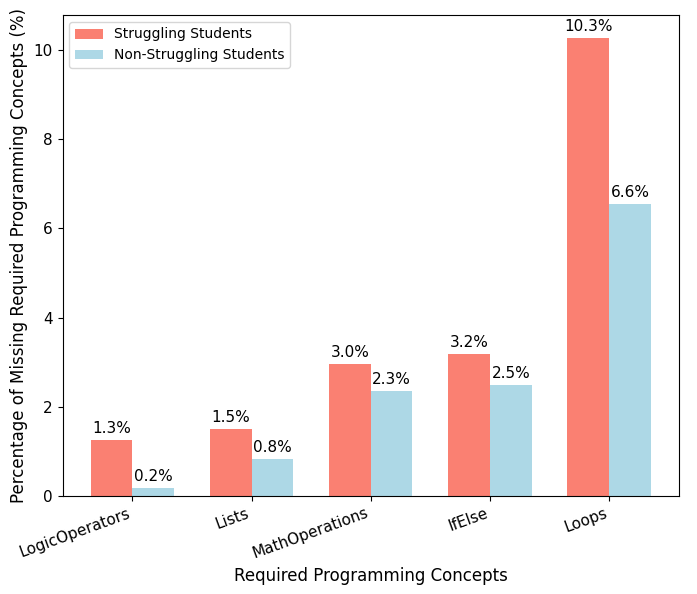}}
\caption{Percentage of lacking programming concepts in CodeWorkout}
\label{fig:MissProgramConstructs}
\end{figure}

Figure~\ref{fig:MissProgramConstructs} shows 
the percentage of lacking programming concepts for key concepts required in CodeWorkout solutions (for both struggling and non-struggling students). 
We determined which programming concepts were lacking by searching students' submissions for code that used the concepts. As shown in the figure, struggling students are more likely to lack programming concepts than non-struggling students, particularly in critical areas such as loops and if-else statements. This underscores the importance of integrating programming concepts into student modeling, as it can enhance the predictive accuracy of identifying struggling students.

\section{Coding Proficiency Taxonomy}
\label{sec:CodingProficiencyTaxonomy}
We conducted in-depth interviews with teachers and teaching assistants (TAs), gathering their own perspectives about struggling student programmers. These interviews informed the construction of a taxonomy of coding proficiencies that students need to apply to complete coding tasks. 

We interviewed two computer science teachers with more than 30 years of teaching experience, as well as two TAs of a university-level introductory programming course (CS1). 

The TAs and teachers completed a survey focusing on the challenges faced by struggling students in computer science.
When asked to describe a struggling student programmer in their own words, the TAs and teachers referred to different stages in the solution process where students commonly struggle. These included a basic understanding of the task, applying concepts learned in class, designing algorithms, and performing debugging and documentation.

The TAs and teachers were also presented with potential indicators of struggling students (e.g., misconceptions, debugging time, final exam grade) and asked to rate how closely these aligned with their own experiences. The characteristics were derived from the literature \cite{alzahrani2018python, edwards2019can,
gorson2020cs1, tabarsicatch, tek2018implicit}, analysis of the datasets' metadata, and common mistakes observed in computer science education. 

All participants agreed that incorrectly applying programming concepts is a key indicator of a struggling student. 

Other indicators were the time required to solve a task, the grade received, and loss of motivation.
The TAs indicated that students who took more than 2.5 times the expected completion time were likely struggling.
 
\begin{figure}
\centerline{\includegraphics[width=9cm]{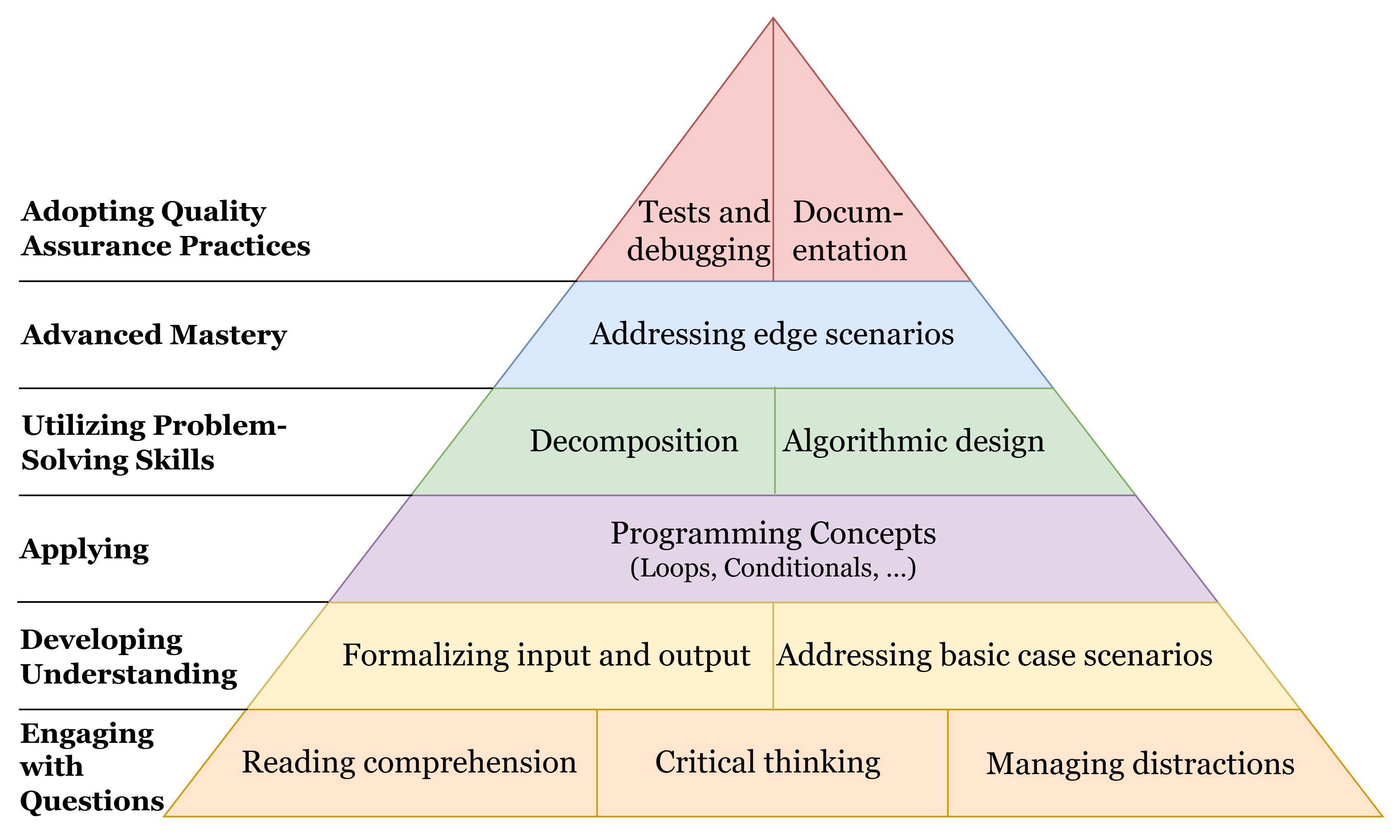}}
\caption{Coding Proficiency Taxonomy: educational goals (underlined, left) are aligned with coding skills (right) }
\label{fig:taxonomy}
\end{figure}

We used the insights collected from teachers and TAs to construct a hierarchy of coding skills that can inform the detection of struggling students with new coding tasks. This hierarchy, called the Coding Proficiency Taxonomy, is
shown in Figure~\ref{fig:taxonomy}, and will be integrated into the detection model described in the next section.

The taxonomy describes levels of educational goals and aligns them with specific coding skills (both observed and latent). It combines insights from the interviews with Bloom's Taxonomy and its variations used for programming~\cite{selby2015relationships, bloom1956taxonomy, thompson2008bloom}. 

As shown in the Figure, the bottom layer of the taxonomy describes how students \emph{engage with the coding task}.  The skills for this include reading comprehension \cite{izu2019fostering}, critical thinking and managing distractions. Teachers emphasized that students must comprehend the coding task text, approach problem-solving independently rather than copying (critical thinking), and manage distractions to remain focused and complete the task. 

The next level up develops \emph{understanding of the task}, such as comprehending inputs and outputs and case scenarios.
The next level focuses on \emph{Applying programming concepts}, such as loops, conditionals, and logical operators. 
Above these is the level \emph{utilizing problem-solving skills},  which includes decomposition and algorithmic design \cite{futschek2006algorithmic}. 
The highest levels of the taxonomy describe \emph{addressing edge scenarios} and \emph{adopting quality assurance practices} such as test cases, debugging and documentation. 

\section{Predicting Struggling Students}
We define the struggling student prediction problem as follows: Given a student's history of submissions for past coding tasks, can we predict whether or not the student will struggle on a new task?

\subsection{The PTM-Model}
The Proficiency Taxonomy Model (PTM) is a multi-task model that simultaneously learns two objectives:
(1)  the student's coding proficiency given their prior task submissions; (2) whether the student is likely to struggle with the new task. 

The model leverages multiple levels of the proficiency taxonomy of Figure~\ref{fig:taxonomy} to predict whether a student is struggling with a new coding task.

\paragraph{Model Input}
The student's history $H$ of $K$ coding tasks is represented as a sequence
 \( H = \{H_1, H_2, \dots, H_{k}\} \),
where each \( H_i \) corresponds to the student's submissions for task \( i \). Each \( H_i \) is itself a sequence \( H_i = \{c_1, c_2, \dots, c_T\} \), with \( c_j \) representing the student's code submission on their \( j \)-th attempt for coding task \( i \), and \( T \) denoting the total number of attempts the student made for task \( i \). 
For each coding task $i$, we also include a struggling indicator $s_i$, which is a binary value assigned at the task level, indicating whether the student struggled on task $i$.
The input to the model consists of the history $H$, the struggling indicators $\{s_1, s_2 \dots, s_k\}$, the target student ID, and the target coding task (including text and programming concepts). 

\begin{figure}
\centerline{\includegraphics[width=6cm]{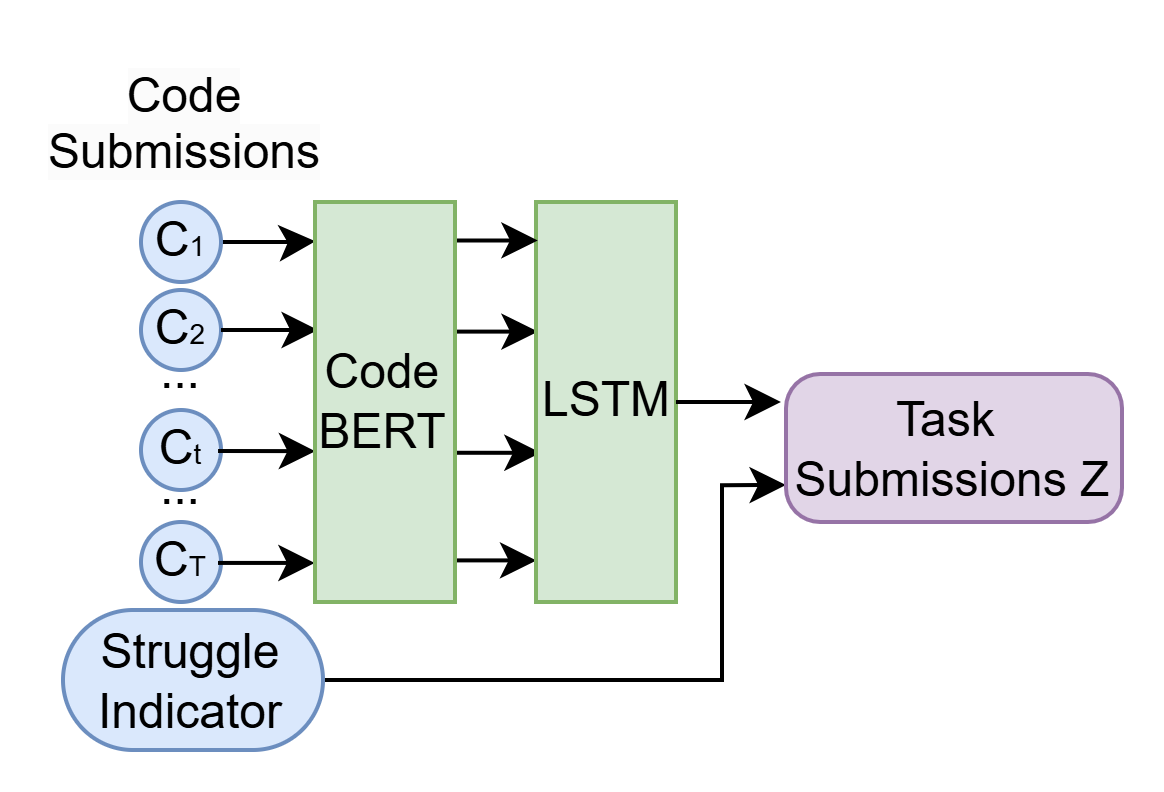}}
\caption{Task Submissions}
\label{fig:task_representation}
\end{figure}

\paragraph{Representing Task Submissions}
The modeling of code submissions of a student for one given task is illustrated in Figure~\ref{fig:task_representation}.
Following past work showcasing the advantages of deep embeddings of students' code~\cite{shi2022code, tsabari2023predicting}, each code submission $c_j$
is embedded using CodeBERT \cite{feng2020codebert}.
These embeddings are processed by an LSTM, which produces a vector $Z$ that represents the student's prior submissions as well as the struggling indicator for that task. 

\subsection{The PTM Architecture}
\begin{figure}
\centerline{\includegraphics[width=9cm]{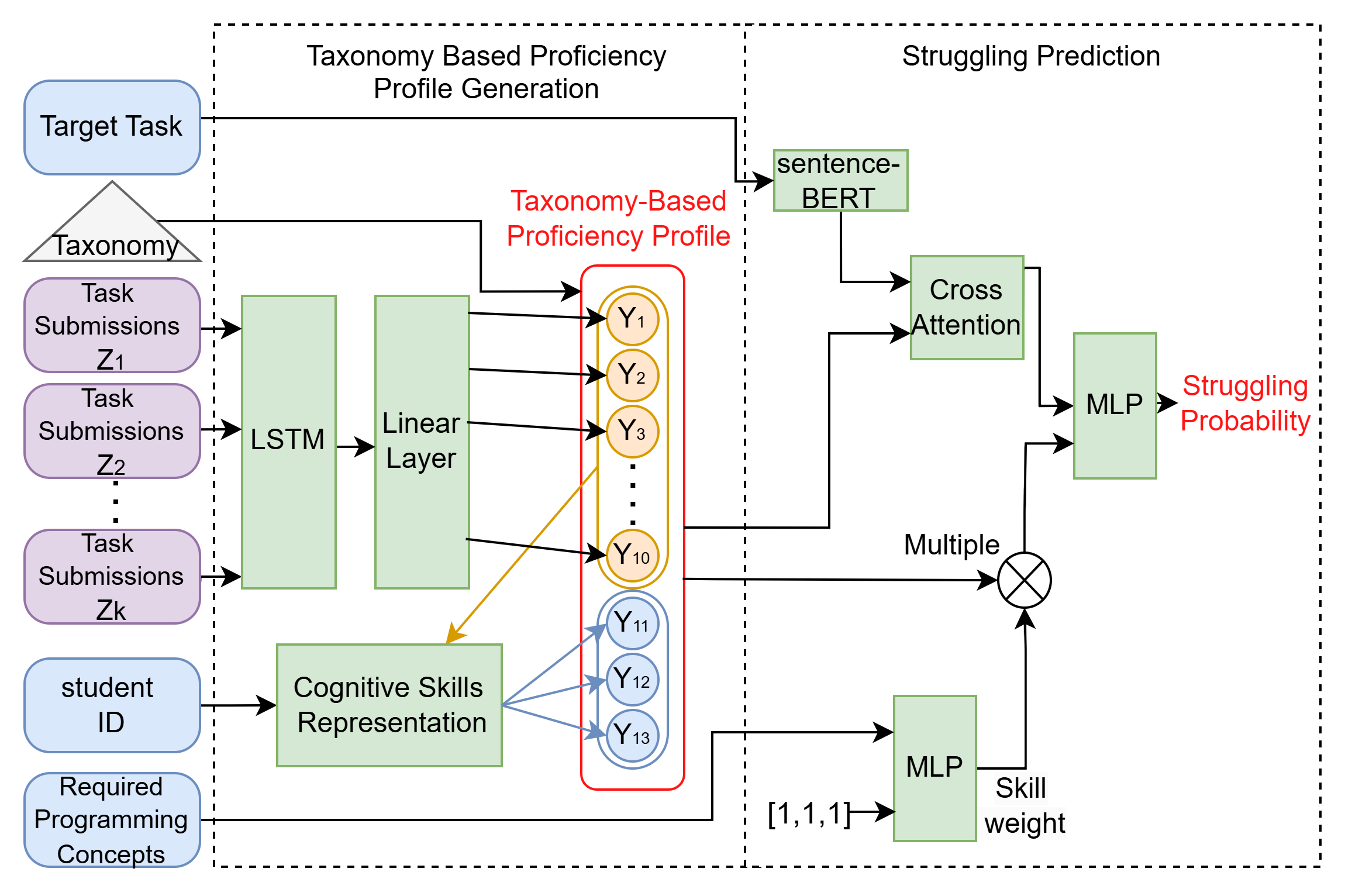}}
\caption{PTM Architecture. Input to the model are task submissions (in purple), student ID, and task information (in blue); outputs from the model are the Taxonomy Based Profile (TBPP), and the struggling indicator (in red); internal model components shown (in green).}
\label{fig:model_architecture}
\end{figure}

The PTM architecture, shown in Figure~\ref{fig:model_architecture}, 
 is divided into two interacting components. The first component generates a Taxonomy-Based Proficiency Profile (TBPP), which is an embedding of the student's proficiency across a subset of skills from the taxonomy.

The TBPP is a vector representing a student's proficiency across 13 dimensions: ten observed skills that explicitly correspond to the taxonomy (i.e., the programming components present in task descriptions), and three latent skills hypothesized to capture additional, implicit aspects of the taxonomy. These latent skills are learned representations that are influenced by both the student ID and the observed skills. The number of latent skills is a tunable hyperparameter chosen during model design.

Each student's skill level is a score between 0 (complete lack of skill) and 1 (full knowledge of the skill). 
An example of one of a student's TBPP in the empirical study is shown below:
\[ TBPP = \begin{bmatrix} \text{Conditional Clauses} \\ \text{Loops} \\ \text{Math Operations} \\ \text{Logic Operations} \\ \text{String Manipulations} \\ \text{Lists} \\ \text{File Operations} \\ \text{Functions} \\ \text{Dictionary}\\ \text{Tuple} \\ \textcolor{black}{\text{L1}} \\
\textcolor{black}{\text{L2}} \\
\textcolor{black}{\text{L3}} \\\end{bmatrix} = \begin{bmatrix} 0.26 \\0.86 \\0.89 \\0.91 \\ 0.93 \\0.43 \\0.45 \\0.29 \\0.43 \\ 0.41\\\textcolor{black}{0.55}\\ \textcolor{black}{0.68}\\ \textcolor{black}{0.5} \end{bmatrix} \]

The student's history of $K$ prior task submissions $(Z_1,\ldots Z_K)$ are passed through an LSTM to produce a single vector representation. This vector is then passed through a linear layer to produce score representations for the student's skills in the 10 possible programming constructs, and these are concatenated with the student's unique ID to create the score representations for the 3 latent skills. This structure reflects the idea that latent skills are unique to each student and inherently linked to their explicit skills. 

The second component uses the student's TBPP to predict whether the student is likely to struggle with a new coding task. 
It combines the student's TBPP with task-specific features (the task text and the required programming concepts) to assess their likelihood of struggling. 
The text of the coding task is encoded using BERT~\cite{devlin2018bert} to generate a contextual representation. A cross-attention mechanism \cite{vaswani2017attention} is applied between the student's TBPP and the BERT representation of the task, where the TBPP is used as the key and query, and the task text is used as the value.
This mechanism captures the relationship between the student's TBPP and the task requirements.

A ``skill weight'' is generated by processing a 13-element binary vector through a Multi-Layer Perceptron (MLP). In this binary vector, each element represents a skill required for the target coding task, with '1' indicating necessity and '0' indicating otherwise. 
The first ten elements of the vector are the skills from the programming concepts for the target task. The final three elements of the vector are the three latent skills. The three latent skills are always considered essential and are represented by three '1's appended to the vector. 
A higher value of the `skill weight' indicates that the task demands a deeper understanding of these components and skills.

The ``skill weight'' is then multiplied by the TBPP, resulting in a weighted TBPP vector. This weighted TBPP vector is concatenated with the output from the cross-attention layer, forming the input to a second MLP. 

The second MLP processes this combined input to produce a binary output, predicting whether the student is struggling with the task.

\subsection{Loss Functions}
The PTM model uses a separate loss function for the TBPP learning task and the struggling prediction task. The combined loss function is defined as:

\begin{equation}
L = a \cdot MAE(\hat{y}, y) + (1 - a) \cdot BCE (\hat{t}, t)
\label{eq:loss_function}
\end{equation}

where $a$ is a weighting factor, and 
\begin{itemize}
    \item \( MAE(\hat{y}, y) \) is the Mean Absolute Error between predicted and estimated TBPP, chosen for its robustness to outliers.

    \item \( BCE(\hat{t}, t) \) is the Binary Cross-Entropy between predicted and true struggling labels, standard for binary classification.

\end{itemize}

The TBPP ground truth for a student is not observed. 
However, it can be estimated as follows:
each student's score for a task is derived from the number of tests passed by the student's last submission. This score is automatically computed in both systems and is available in the datasets. Let $S$ denote the student's score on previous $K$ tasks (a matrix of dimensions $(K,1)$). 
Let $A$ denote the binary vector representation of the programming concepts and skills required for the $K$ tasks (a matrix of dimensions $(13, K)$).
The ground truth estimation of the TBPP is the multiplication of $A$ and $S$. 
An example is shown below:
\[
A \cdot S = \begin{bmatrix}
1 & 0 & 1 & \dots & 1 \\
0 & 1 & 0 & \dots & 1 \\
1 & 0 & 1 & \dots & 1 \\
\vdots & \vdots & \vdots & \ddots & \vdots \\
1 & 1 & 0 & \dots & 1
\end{bmatrix}
\cdot
\begin{bmatrix}
s_1 \\
s_2 \\
s_3 \\
\vdots \\
s_{k}
\end{bmatrix}
=
\begin{bmatrix}
y_1 \\
y_2 \\
y_3 \\
\vdots \\
y_{13}
\end{bmatrix}_{13 \times 1},
\]

In the resulting vector, each \( y_i \) represents the weighted contribution of final scores to the \( i \)-th component, summarizing the student's proficiency in that component.
To ensure consistency, we normalize the TBPP estimation by scaling it based on the range of scores across all tasks. This prevents any skill from being overrepresented.

\section{Experiments}
We evaluate the PTM model against baseline approaches in an extensive set of experiments. We follow the same empirical methodology as in the CSEDM Data Challenge~\cite{csedm2021competition}. Specifically, the first 30 task submissions of the student are used to predict whether the student will struggle in the following 20 submissions. By designing the task in this manner, the accuracy of the model is tested without retraining after each new submission, reflecting a realistic class scenario while preserving the temporal consistency of the data.

Our experiments aim to
address the research questions outlined in Section~\ref{sec:research}
by examining the effectiveness of deep language models for software code representation (RQ1), the role of prior coding
snapshots and submission history (RQ2), and the contribution of a Coding Proficiency Taxonomy in predicting struggling students (RQ3).
We compare the performance of our approach against several baselines:
\paragraph*{DKT with Target Task ID}

This baseline builds on the DKT model of Piech et al.~\cite{piech2015deep}, incorporating the following modifications: the target task ID was embedded and appended to the final hidden layer of the LSTM, providing a representation of the target task for prediction. Additionally, the correctness label, which indicates whether the student answered correctly, was replaced with a struggling label indicating whether a student is struggling with the target task.
\paragraph*{Code-DKT with Target Task ID}

The Code-DKT model \cite{shi2022code} extends DKT by integrating code representations: tasks are one-hot encoded, and code attempts use ASTs processed via code2vec. As the original repository\footnote{\url{https://github.com/YangAzure/Code-DKT}} only supports Java, we added AST support for Python, which is included in our released code. As with DKT, we replaced correctness labels with struggling labels and concatenated the final hidden state with the target task ID (one-hot encoded) before applying a linear layer. The original loss function was preserved.
\paragraph{SAKT}
This baseline uses the SAKT model of Pandey and Karypis~\cite{pandey2019self}, which applies a self-attention mechanism to model temporal interactions between past tasks and the current prediction. The model incorporates the target task via a query embedding derived from the final interaction. We used the classical SAKT implementation available in a public GitHub repository~\footnote{\url{https://github.com/seewoo5/KT}}.

\paragraph{Experimental Setup}
\label{sec:Experimental Setup}
In all experiments, we applied a 5-fold cross-validation strategy (CV-5). The dataset was split on a per-student basis, ensuring that each student appeared exclusively in either the training or testing sets. 
For each dataset, we trained a separate model, resulting in one model for CodeWorkout and another for FalconCode to account for their distinct characteristics.

For each coding task, we included up to 100 recent attempts. 
We employed the loss function that equally balances the MAE for the TBPP regression and the BCE loss for the struggling label classification (assigning 0.5 to $\alpha$ in Equation~\ref{eq:loss_function}).

All LSTMs in our model were configured with a hidden layer size of $512$. The models were trained for 15 epochs using a fixed learning rate of 0.0001 and a batch size of 32. These settings were chosen based on hyperparameter recommendations from the literature~\cite{bengio2012practical}.

\section{Results}
We now present the experimental findings focusing on four main areas: (1) comparisons with baseline models, (2) ablation studies, (3) sensitivity analysis, and (4) performance analysis across datasets.

To evaluate model performance, we use the ROC-AUC metric~\cite{huang2005using}, which measures the model's ability to distinguish between struggling and non-struggling students by assessing its classification performance across various thresholds. 
ROC-AUC is a standard evaluation metric in educational settings~\cite{piech2015deep,bowers2019receiver,shi2022code} and is particularly useful for imbalanced datasets.

Statistical significance tests were performed for all results using the paired bootstrap test~\cite{thompson1993use,effron1993introduction,bisani2004bootstrap}. The paired bootstrap approach is particularly advantageous since it empirically estimates variability and confidence intervals without relying heavily on distributional assumptions, making it robust even when traditional assumptions (e.g., normality) may not hold. As Thompson~\cite{thompson1993use} highlights, it provides a practical way to assess the reliability and stability of findings by generating numerous resampled datasets, enhancing the interpretability of significance tests. Bonferroni corrections~\cite{bonferroni1936teoria} were performed to compensate for multiple hypothesis testing.
A star mark (``*'') in the results tables (tables \ref{tab:perfcom}, \ref{tab:ablation}) denotes a model is significantly better than the rest.

\subsection{Comparisons with Baseline Models}
 Table \ref{tab:perfcom} presents a comparison with baseline models. As seen in the table, PTM outperformed all other methods on both datasets. We list several factors that may have contributed to the differences in performance. 

First, both DKT and Code-DKT models represent coding tasks using one-hot encoding, which may limit their ability to capture nuanced relationships between tasks, potentially affecting prediction accuracy.

Second, the original Code-DKT model was designed to handle a limited number of coding tasks (e.g., 10 tasks). In contrast, our current datasets, CodeWorkout and FalconCode, include approximately 50 and 400 tasks respectively. This significant increase in task variety and sequence length may have challenged Code-DKT's capacity to process extended sequences effectively.

Furthermore, the SAKT model achieved strong results in both data sets, highlighting its strength in modeling temporal dependencies in student interactions. Nevertheless, PTM consistently outperformed SAKT, indicating that PTM captures more complex patterns in student learning behaviors.

These findings highlight the strengths of our model in managing complex task representations and extended histories.

\begin{table}[ht]
\centering
\caption{Performance Comparison}
\begin{tabular}{|l|c|c|}
\hline
\multicolumn{2}{|c|}{\textbf{Results for CodeWorkout Dataset}} \\ 
\hline
\textbf{Model} & \textbf{\shortstack{\small{\textit{ROC-AUC}}[\%]}} \\ 
\hline
DKT with Target Task ID & $76.27 \pm 2.2$ \\ 
Code-DKT with Target Task ID & $70.93 \pm 4.3$ \\ 
SAKT & $76.11 \pm 1.09$ \\
PTM & \textbf{\boldmath$77.09 \pm 1.8$*} \\ 
\hline
\multicolumn{2}{|c|}{\textbf{Results for FalconCode Dataset}} \\ 
\hline
\textbf{Model} & \textbf{\shortstack{\small{\textit{ROC-AUC}} [\%]}} \\ 
\hline
DKT with Target Task ID & $63.54 \pm 1.8$ \\ 
Code-DKT with Target Task ID & $ 62.17 \pm 2$ \\
SAKT & $70.52 \pm 0.8$ \\
PTM & \textbf{\boldmath$73.18 \pm 0.9$*} \\ 
\hline
\end{tabular}
\label{tab:perfcom}
\end{table}

\subsection{Ablation Studies}
The ablation studies evaluated the contributions of the coding proficiency taxonomy and historical coding data to predict whether a student is struggling. To do this, we modified the PTM model by systematically removing components and analyzing the impact on performance.

We used the following modified single-task models to predict a binary struggling outcome:

\begin{itemize}
    \item \textbf{No-Tax:} Excludes the taxonomy features but retains all task information (including past submission attempts).
    \item \textbf{No-Tax \& No-Hist:} Excludes taxonomy features and uses only the last submission information for each task.
\end{itemize}

Each model follows a simplified struggling prediction component, where the target task (processed via BERT) and the 10 observed skills (from the TBPP) pass through a separate MLP, combined via weighted addition (weights learned during training), and then processed through a linear layer for the final prediction.

Table \ref{tab:PerformanceComparison} summarizes the ablation results, obtained using the same empirical configuration as in Section \ref{sec:Experimental Setup}. As seen in the table, PTM outperforms both ablation conditions on both datasets. 

The results also show an improvement as we progressively add more information to the model across both datasets. The lowest performance is observed in the No-Tax \& No-Hist models, where neither history nor taxonomy are included. As we add only the student's coding history (in the No-Tax condition) results improve. Further incorporating the proficiency taxonomy leads to the PTM model, which achieves the best performance. 

\begin{table}[h]
\centering
\caption{Ablation studies on both datasets}
\label{tab:PerformanceComparison}
\begin{tabular}{|l|c|c|}
\hline
\multicolumn{2}{|c|}{\textbf{Results for CodeWorkout Dataset}} \\ 
\hline
\textbf{Model} & \textbf{\small{\textit{ROC-AUC}} [\%]} \\ 
\hline
No-Tax \& No-Hist & $72.50 \pm 3.5$\\ 
No-Tax & $75.52 \pm 1.7$ \\ 
PTM & \textbf{\boldmath$77.09 \pm 1.8$*} \\ 
\hline
\multicolumn{2}{|c|}{\textbf{Results for FalconCode Dataset}} \\ 
\hline
\textbf{Model} & \textbf{\shortstack{\small{\textit{ROC-AUC}} [\%]}} \\ 
\hline
No-Tax \& No-Hist & $69.13 \pm 1.3$ \\ 
No-Tax & $71.33 \pm 0.6$ \\ 
PTM & \textbf{\boldmath$73.18 \pm 0.9$*} \\ 
\hline
\end{tabular}
\label{tab:ablation}
\end{table}

\subsection{Sensitivity Analysis}
\begin{figure}[t]
\centerline{\includegraphics[width=8cm]{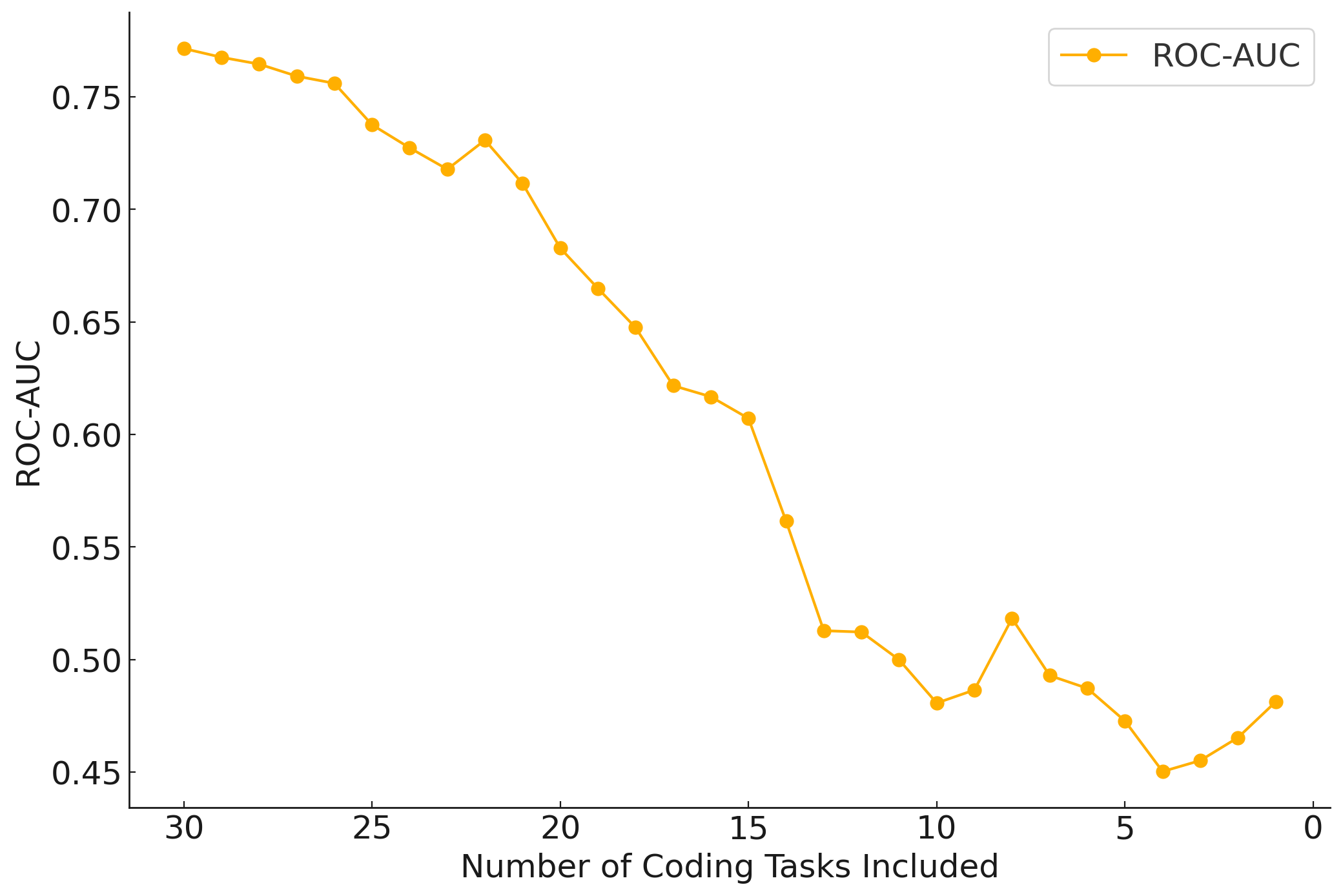}}
\caption{Sensitivity of ROC-AUC to the number of coding tasks included in student Histories (CodeWorkout Dataset)}
\label{fig:Sensitivity}
\end{figure}

We analyzed the sensitivity of PTM to varying lengths of student coding tasks, assessing its ability to make accurate predictions with different amounts of past-task data.
Figure \ref{fig:Sensitivity} explores PTM's sensitivity to different lengths of student past tasks solved in the CodeWorkout dataset. The FalconCode results are similar.

For each point, we vary the number of coding tasks included during inference and calculate prediction success on future tasks. Starting with 30 tasks, we progressively remove the earliest tasks, until only the most recent coding task remains in the input. 

The figure illustrates how performance improves when more past tasks are included, highlighting the importance of extended input sequences for PTM. In general, the model effectively captures patterns in student performance when sufficient prior data is available.
Between $22$ and $30$ past tasks, the ROC-AUC remains relatively stable. Once the number of included tasks drops further, we observe a steep decline in ROC-AUC.

\subsection{Result Differences Across Datasets}
We observed a statistically significant drop in PTM's performance on the FalconCode dataset compared to the CodeWorkout dataset. To understand this discrepancy, we analyzed the tasks assigned to students in both systems and identified key differences that may account for the performance gap:

\paragraph{Task Structure and Complexity} The CodeWorkout dataset features uniformly structured tasks with similar difficulty and response length, possibly simplifying modeling. In contrast, FalconCode includes both short fundamental exercises and longer, more complex lab tasks. This variability increases modeling difficulty due to the broader range of coding styles and solution complexities.
\paragraph{Task Ordering} CodeWorkout tasks follow a fixed sequence, giving students a similar task order and reducing variability, which potentially aids model learning. FalconCode offers a diverse task pool with varying sequences, increasing variability and potentially reducing model performance. 

\paragraph{Locally Solved Coding Tasks} In FalconCode, students may compile code locally and submit only when seeking feedback. Snapshots of local work exist, but compilation outcomes are missing. We excluded tasks explicitly marked as local, though unmarked cases may remain. This partial visibility limits modeling because the removal of locally completed tasks creates gaps in the students' coding histories, making the task order incomplete. In contrast, CodeWorkout tasks are fully online, enabling complete and consistent data.

\section{Discussion and Limitations}
Our study demonstrates the power of the PTM model in predicting struggling students on multiple datasets by (1) leveraging deep learning models (specifically CodeBERT) to capture structural and semantic patterns in student code (RQ1), (2) incorporating prior code snapshots to enhance predictions by providing a view into learning progress (RQ2), and (3) representing students' coding proficiency, to account for required and demonstrated student skills (RQ3).
These findings highlight the importance of both long-term coding processes and taxonomy-based features in capturing nuanced differences in student understanding, as evidenced by consistent improvements in ROC-AUC results across both datasets. We discuss multiple considerations and highlight several limitations of our work.

First, as programming education adapts to the generative AI era, the type and scope of coding tasks are shifting. These tasks are progressively increasing in complexity, evolving from fundamental concepts like variables, loops, conditions and functions to more advanced topics and to larger assignments~\cite{guner2025ai}. Despite these changes, we believe that core programming skills, including those represented in our proposed taxonomy, remain essential~\cite{prather2023robots, guner2025ai}. These enduring skills highlight the continued relevance of our research in supporting student learning and in predicting their struggle points.

Second, identifying genuine struggle remains a challenge, as students might appear successful while relying on AI tools or copying from their peers instead of developing their own problem-solving skills. 
Specifically, students now have access to instant, often correct, GenAI-based solutions to programming tasks~\cite{prather2023robots} that require minimal effort on their part. This phenomenon was not considered in this work and is left for future research.

Third, we tested CodeT5 and CodeLlama as alternatives to CodeBERT. Both showed small gains, but at higher computational cost. We also fine-tuned QWen as an end-to-end baseline, but it underperformed, likely due to weak supervision and the temporal nature of the task.

Fourth, the unit tests used as measures in our empirical evaluations focus on functional correctness, but may not fully capture learning or conceptual understanding. For example, CodeWorkout includes hidden tests that check edge cases, yet passing them doesn't always reflect deep comprehension.

Fifth, while our primary evaluation metric was ROC-AUC, we did not include precision or recall scores due to the imbalanced nature of the datasets. We also computed the calibration of predicted probabilities \cite{guo2017calibration}, which assesses how well the model's confidence aligns with actual outcomes. PTM and SAKT demonstrated the best calibration.

Finally, not all elements of the proficiency taxonomy were explicitly modeled in PTM. We included three latent skills to capture abstract processes and used explicit programming concepts to reflect observable behaviors. Skills like edge case handling and basic understanding were indirectly captured via score-based loss. Some layers were excluded due to dataset limitations, for example, testing was omitted as unit tests were provided, and documentation was not required. In future work, we plan to extend the taxonomy coverage.

 \section{Conclusion and Future Work}
 In this study we present a new model, called Proficiency Taxonomy Model (PTM), for detecting struggling student programmers given their prior coding history. By leveraging coding proficiency taxonomies and language models, PTM outperforms the existing state-of-the-art models across datasets and programming languages. 
 
 Our findings emphasize the importance of combining pedagogical insights with language models. The inclusion of a structured coding proficiency taxonomy ensures that predictions align with educators' understanding of student difficulties, allowing for more targeted interventions. Furthermore, our results indicate that capturing the entire coding history of a student improves predictive performance, reinforcing the value of long-term tracking in educational modeling.
As software programming education evolves, incorporating adaptive driven support and refining taxonomy representations will be crucial in addressing emerging challenges, including AI-assisted coding and shifting educational paradigms. 
 
Future work will aim to broaden our model's scope to capture a wider range of programming competencies and adapt to evolving student skills.
Another possible direction would be developing a unified cross-domain model trained simultaneously on both the Java and Python datasets. Such a unified model could potentially identify language-independent indicators of student struggle while accounting for language-specific characteristics. 

Finally, the current estimation of students' TBPP gives equal weight to all tasks in their history. Future work could investigate weighting tasks based on factors such as recency or difficulty, potentially leading to a more accurate and dynamic representation of student proficiency.

Ultimately, this research could lead to the development of targeted educational tools that empower educators and enhance the support they are able to provide to students who are learning fundamental programming skills. This is especially important in large classrooms where students' prior coding expertise is diverse. To this end we intend to deploy our tool in classrooms and study its practical applications.



\bibliography{sigproc}

\end{document}